\documentclass[conference,a4paper]{IEEEtran}

\usepackage[a4paper,bindingoffset=0.2in,%
            left=0.507874in,right=0.517874in,top=1in,bottom=1in,%
            footskip=.25in]{geometry}
            
\usepackage{amsmath}
\usepackage{url}
\usepackage{tabto}
\usepackage{xparse}
\usepackage{cite}
\usepackage{graphicx}
\usepackage{makecell}
\usepackage{enumerate}
\usepackage[font=footnotesize,skip=0.1pt]{caption}

\begin{document}
\title{Insights from Analysis of Video Streaming Data to Improve Resource Management}

\author{\IEEEauthorblockN{Sabidur Rahman\IEEEauthorrefmark{1},
Hyunsu Mun\IEEEauthorrefmark{2},
Hyongjin Lee\IEEEauthorrefmark{2},\\
Youngseok Lee\IEEEauthorrefmark{1}\IEEEauthorrefmark{2},
Massimo Tornatore\IEEEauthorrefmark{1}\IEEEauthorrefmark{3}, and
Biswanath Mukherjee\IEEEauthorrefmark{1}}
\IEEEauthorblockA{\IEEEauthorrefmark{1}University of California, Davis, USA  \IEEEauthorrefmark{2}Chungnam National University, Korea \IEEEauthorrefmark{3}Politecnico di Milano, Italy}
\IEEEauthorblockA{Email: \{krahman, mtornatore, bmukherjee\}@ucdavis.edu, \{munhyunsu, hjlee201203399, lee\}@cnu.ac.kr}}

\IEEEoverridecommandlockouts
\IEEEpubid{\makebox[\columnwidth]{ 978-1-5386-6831-3/18/\$31.00~\copyright2018 IEEE\hfill} \hspace{\columnsep}\makebox[\columnwidth]{ }}
\maketitle
\IEEEpubidadjcol

\begin{abstract}
Today a large portion of Internet traffic is video. Over The Top (OTT) service providers offer video streaming services by creating a large distributed cloud network on top of a physical infrastructure owned by multiple entities. Our study explores insights from video streaming activity by analyzing data collected from Korea's largest OTT service provider. Our analysis of nationwide data shows interesting characteristics of video streaming such as correlation between user profile information (e.g., age, sex) and viewing habits, viewing habits of users (when do the users watch? using which devices?), viewing patterns (early leaving viewer vs. steady viewer), etc. Video on Demand (VoD) streaming involves costly (and often limited) compute, storage, and network resources. Findings from our study will be beneficial for OTTs, Content Delivery Networks (CDNs), Internet Service Providers (ISPs), and Carrier Network Operators, to improve their resource allocation and management techniques.
\end{abstract}


\section{Introduction}
Internet users are using more and more Video on Demand (VoD) applications such as Netflix, Amazon, YouTube, etc. Even traditional live television channels are now being streamed by service providers such as Sling TV, DirectTV Now, etc. According to Cisco Visual Networking Index (VNI)~\cite{cisco} 80-90\% of the traffic in 2021 will be video.

VoD services require compute, storage, and network resources which are costly and often limited. Hence, new effective resource management methods are required to serve ever increasing video traffic. These methods should be scalable, adaptive, and aware of patterns related to user activity, spatio-temporal variation of load, user profiles, devices used, etc. Hence, understanding these patterns is very important for resource management of carrier networks, DC networks, Content Delivery Networks (CDNs), Internet Service Providers (ISPs), etc.

There have been a few prior works studying partial analysis of VoD services. For example, Ref.~\cite{chen} analyzes user behavior in VoD traffic, specially video watching sessions. The data analyzed in the study was collected before 2014 and uses Hadoop system to process the data. Ref.~\cite{li} investigates characteristics of user behavior in mobile live streaming systems. But the data only accounts for live contents and mobile devices, hence it misses out on large portion of contents and users. Due to privacy concerns, limitation of publicly available data from large video streaming services, this area of study lags behind compared to areas such as analysis of mobile traffic patterns~\cite{fxu} and mobile application usage~\cite{marquez}. To fill this void, our study reports insights from a comprehensive nation-wide dataset including live channels and video contents, traces collected over both wireless and wired network, etc. Another unique feature of the data is that it not only contains user's viewing traces, content types and identification in details, but also contains anonymized information such as user's age group, device IP address, device type, etc.

To the best of our knowledge, our study is the first to use big data analytics tools Apache Spark~\cite{spark} and Zeppelin~\cite{zep} to analyze video streaming activity. The size and scale of data used in our study required more than traditional analytics tools. To keep the analysis scalable for future studies and deployable in practical run-time scenarios, we have used big data analytics tools. The results from our analysis enable us to explore answers for the following questions:
\begin{itemize} 
\item Analysis of user profiles: Who is watching the videos?
\item Temporal patterns in viewing activity: When do users watch video?
\item Content-centric viewing patterns: Which contents are popular? When are popular contents viewed?
\item Video browsing behavior: How can we correctly classify video browsing behaviors (early leaving, steady viewer, etc.)?
\item User device: Which device(s) do users use to watch certain videos?
\end{itemize}

The rest of this study is organized as follows. Section II provides a description of the dataset and methodology. Section III describes how we use big-data analytics to answer questions that are important for infrastructure owners and service providers. Section IV concludes the study and indicates directions for future works.

\section{Dataset And Methodology}
\subsection{Description of Dataset}
Our study uses dataset collected from the largest OTT in Korea. This nationwide data was collected in 2017 and the results in our study are derived from 24-hour data collected from 3M subscribers on a weekday.

The data considers 70 live channels, 7000 movies, and 280,000 other VoD contents. The data trace was collected every 10 seconds. Tables~\ref{view_info} and~\ref{user_info} show sample data with explanation followed.
\begin{table}[!htb]
\caption{Sample data: viewing information.}
\label{view_info}
\centering
\begin{tabular}{|c|c|c|c|c|c|c|c|c|c|c|}
\hline
D&H&M&S&\makecell{U.\\ID}&T.&\makecell{P.\\ID}&\makecell{M.\\T.}&Dev.&BR&IP\\
\hline
d&10&01&10&a3&L&A01&01:20&\makecell{Andr.\\Ph.}&2&p\\
\hline
d&10&01&20&a3&L&A01&01:30&\makecell{Andr.\\Ph.}&2&p\\
\hline
d&10&01&20&a5&V&A01&00:05&\makecell{Smrt.\\TV}&1&q\\
\hline
\end{tabular}
\end{table}

In Table~\ref{view_info}, the first four columns (D, H, M, and S) displays date, hour, minute, seconds of the trace. The fifth column contains user identification (e.g., a3) which helps us to map the viewing info with user profile from Table II. Sixth column (T) reports `the type of video content': `L' for live contents, `V' for non-live video contents. Seventh column (P.ID) contains the content identification.
Eighth column (M.T.) contains the media time inside the content where viewer is watching now. Ninth column (Dev.) contains the device type the viewer is using. We have seven major types of devices in our study data: Android Phone, iOS Phone, Android Tablet, iOS Tablet, PC, Smart TV, Chromecast. Tenth column (BR) contains the `bitrate' requirement for the connection (`2' stands for 2 Kbps and so on). Eleventh column (IP), contains the IP address (e.g., `p') of the customer device with lower 16 bits hidden (for example, `p' = 192.168.*.*).

Table~\ref{user_info} contains user profile information such as gender and age group of the user. This helps us to explore the impact of such information in viewing behavior and more.
\begin{table}[!htb]
\caption{Sample data: user profile information.}
\label{user_info}
\centering
\begin{tabular}{|c|c|c|}
\hline
\makecell{User\\ID} & Gender & \makecell{Age\\Group}\\
\hline
a3 & M & 30\\
\hline
a5 & F & 20\\
\hline
\end{tabular}
\end{table}

\subsection{Methodology and Analysis Environment}
For big data analysis environment, we have used Apache Spark 2.2.0 and Apache Zeppelin 0.7.3. Apache Spark is an advanced analytics engine for large-scale data processing. Apache Spark uses both batch and streaming data to gain high performance. Ref.~\cite{spark} reports hundred times faster performance compared to Hadoop.

Traditional methods such as standard database queries and Hadoop distributed processing can generate these results as well, when there is no computation time constraints. But, when we start applying this results in real-time, tools like Apache Spark are a better option in terms of computation time and large-scale data handling.

\section{Findings and Results: What questions can data answer?}
We study and analyze the data to understand the relationship between user activity and features such as time of the day, users' age, content type, device type, etc. From our findings through data analysis, we report the following results.

\subsection{Analysis of user profiles: Who is watching the videos?}
Table~\ref{agegroup} shows the breakdown of number of viewers (\%) in different age groups. According to Table~\ref{agegroup} maximum number of users come from age group 30 (39.54\%), followed by age group 20 (31.09\%). Such findings can help to understand the distribution of users' age groups.
\begin{table}[!htb]
\caption{Age group analysis for viewers.}
\label{agegroup}
\centering
\begin{tabular}{|c|c|c|}
\hline
\makecell{Age group} & \makecell{Percentage}\\
\hline
Less than 10 & 0.001\\
\hline
10-19 & 2.48\\
\hline
20-29 & 31.09\\
\hline
30-39 & 39.54\\
\hline
40-49 & 17.16\\
\hline
50-59 & 7.09\\
\hline
60-69 & 1.78\\
\hline
Greater than 70 & 0.798\\
\hline
\end{tabular}
\end{table}

The gender breakdown of viewers is 53.54\% female, 42.68\% male and 3.78\% others.

\subsection{Temporal patterns in viewing: When do users watch video?}
\subsubsection{Impact of age}
\begin{figure}[!htb]
\centering
\includegraphics[width=3.5in]{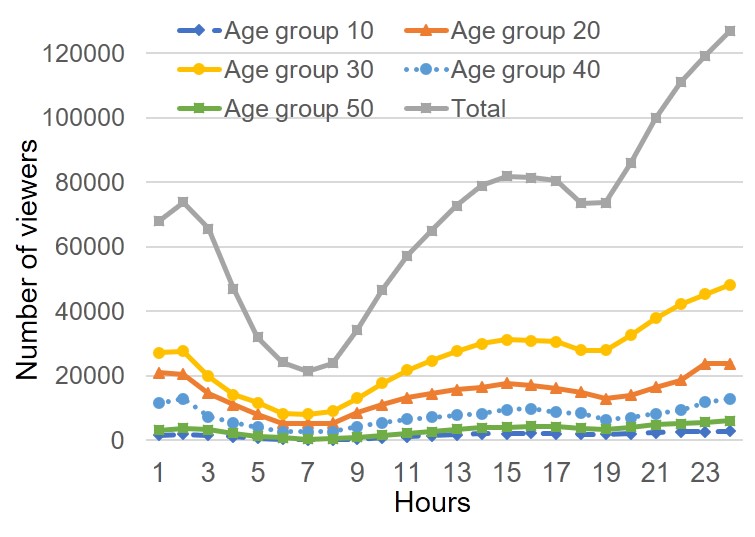}
\caption{Actual number of active viewers over 24 hours, from different age groups.}
\label{active_actl}
\end{figure}

Fig.~\ref{active_actl} shows the actual number of users over 24 hours, distributed among different age groups. We observe that Viewing activity increases as the day progresses, for all age groups. Also, viewing activity increases rapidly after 1900, one explanation of this can be: as more users come back home, number of active viewers grows. In addition, users from age groups 20 and 30 dominate throughout the day. This phenomenon can be explained by Table~\ref{agegroup}, as we have observed the largest number of viewers come from age group 20 and 30.

\subsubsection{Impact of gender}
\begin{figure}[!htb]
\centering
\includegraphics[width=3.5in]{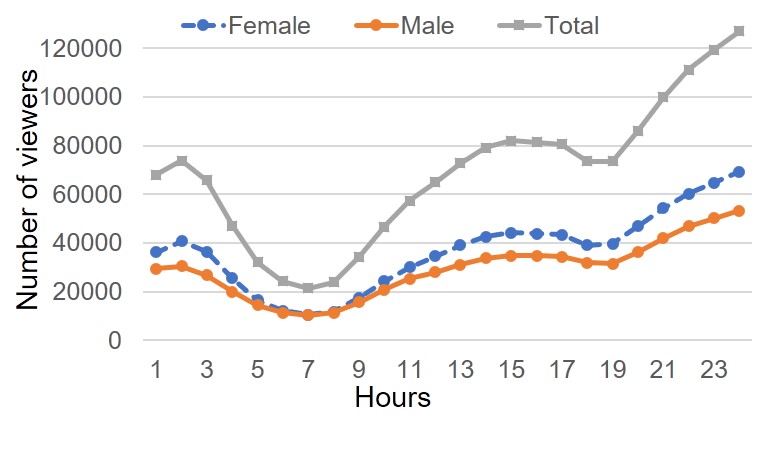}
\caption{Number of active viewers over 24 hours, from different genders.}
\label{view_gender}
\end{figure}

Fig.~\ref{view_gender} shows the hourly user count over 24 hours, distributed among different genders. We observe similar patterns as Fig.~\ref{active_actl}. Viewing activity increases as the day progresses for both male and females. Contrary to popular beliefs, we observe more female viewers throughout the day. This phenomenon, again, can be explained by the fact that larger portion of the users are from female gender.

\subsection{Content-centric viewing patterns: Which contents are popular? When are popular contents viewed?}
\begin{figure}[!htb]
\centering
\includegraphics[width=3.5in]{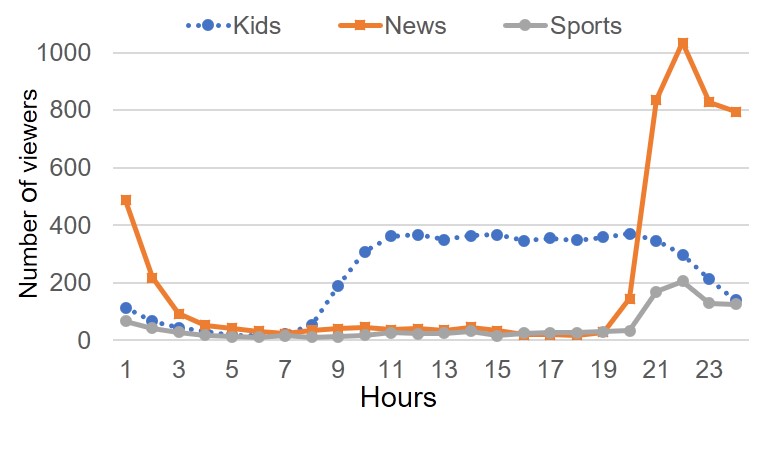}
\caption{Viewing patterns for different types of contents, over 24 hours.}
\label{sport-news}
\end{figure}

Fig.~\ref{sport-news} shows very interesting pattern in three different (types) of contents. Content relevant for ``Kids'' is popular most of the active day-time and late at night. On the other hand, news related content is popular only at night (1900-2300) as the adult age group are actively viewing during that period. Similar pattern is also followed by ``Sports''. Content placement and network management methods unaware of such trends might make inaccurate decisions, leading to QoS violations and additional operational cost.

\subsection{Video browsing behavior: How can we correctly classify video browsing behaviors?}
One of the most interesting insights of our study is the viewer's behavior during the viewing. After studying the data, we have classified the viewer's viewing pattern in the following four categories:
\begin{itemize}
\item Early leaving: stops watching content in 5 minutes.~\footnote{\label{viwer_footnote}Prior study~\cite{chen} examines the early-leaving viewer behavior and shows why `5 minutes' is a practical threshold.}
\item Steady viewer: watches a single video for more than 5 minutes.
\item Highlighter: browses only the interesting parts of a video.
\item Surfing watcher: watches several videos within an hour.
\end{itemize}

Our data indicates that the viewers do not usually watch the whole content. A large portion of them watches only the interesting parts (highlighters are 56.44\%), followed by the \emph{steady viewers} (25.74\%). In addition, there is a significant number of \emph{early leaving viewers} (10.89\%) too, who leaves the system in less than 5 minutes. \emph{Surfing viewers}, who watch multiple contents within an hour consists of 7.0\%. Hence, resource management methods should prepare the system to serve \emph{highlighters} and \emph{surfing watchers}, in addition to \emph{steady} and \emph{early leaving viewers}.

\subsection{User device: Which device(s) do users use to watch videos?}
From our analysis, the breakdown of the devices used for video streaming is as follows: `Android Phone' leads with 33.5\%, followed by `PC' 29.5\% and `iOS Phone' 16.6\%. Rest of the devices are `iOS Tablet' (9.6\%), `Android Tablet' (5.1\%), `Smart TV' (5\%), and others (0.7\%). Content quality and content size (4K HD vs. 1080P) will change with the device and screen size making the device type is an important parameter in content and network management problems.

\section{Conclusion}
Our study presents an analysis of nationwide data collected from Korea's biggest OTT service provider. We use big data analytics tools (Apache Spark and Zeppelin) to analyze the data. Our analysis shows interesting insights into user behavior and helps us to look for answers of important questions related to video streaming. Future studies can use such understandings to develop network and content management methods which are more data-driven. Predicting future user behavior (steady viewer vs. early leaving) by learning from previous patterns is one interesting direction to explore.

\textbf{Acknowledgement}

Youngseok Lee was supported by Basic Science Research Program through the NRF of Korea funded by the Ministry of Education (NRF-2016R1D1A1A09916326) and by the MSIT of Korea under the ITRC support program (IITP-2018-2016-0-00304).

\end{document}